# New collective zero-resistance states in GaAs/AlGaAs heterostructures

J. C. Phillips

*Dept. of Physics and Astronomy, Rutgers University, Piscataway, N. J., 08854-8019*

## Abstract

Exponentially small resistance in crossed static magnetic and microwave fields occurs in ultrapure two dimensional electron gases (2DEG), while in less pure systems the magnetoresistance oscillates but is always positive. A non-perturbative theory of self-organization explains the new collective states in terms of open orbits. There are analogies to many other anomalous physical phenomena, as well as to scaling and evolutionary principles of biophysics.

Zudov *et al.* discovered[1] giant magnetoresistance oscillations associated with low-field cyclotron resonance in samples previously used to study quantum Hall effects; recently Mani *et al.*[2] and they[3] have shown that the minima of these oscillations correspond to exponentially small resistance. A number of microscopic models have been proposed, including some kind of sliding charge density wave[1,4] and pseudobinding of electrons near the Fermi surface via excitons in a way that is reminiscent of the excitonic mechanism of superconductivity[3]. The central theoretical problem is to understand how the incident microwave field (MW) combines with the crossed magnetic field B to generate these new collective states. This is a very complex problem in glassy disorder, and it appears to exhibit many of the characteristics of self-organization, a phenomenon of widespread applicability in non-crystalline (disordered) systems[4].

At first sight it would appear that the crossed MW and B fields cannot generate new states, but the magnitude of the new effects increases dramatically as mobilities and relaxation times increase, by factors of 5 and 8 in refs. 2 and 3, relative to ref. 1. This is similar to high-temperature superconductivity in the cuprates, where correlations with anomalous normal-state transport are observed, but not to conventional superconductivity in normal metals. It is also similar to pinning of charge density waves. This means that the theory cannot rely on first-order dynamics of the crossed fields, but must incorporate the longer time (possibly glassy) dynamics that can dominate transport properties in very high mobility samples.

Mani *et al.*[2] have noted that there is a 1/4-cycle phase shift of the resistance minima with respect to integral cyclotron frequencies with $x^{-1} = B/B_1$, $B_1 = \omega m^*/e$, and $\omega_{cn} = \omega/n$ (or $B_n = B_1/n$), where there are sharp resistance maxima[3]. They explain this shift in terms of an excess configurational energy $\Delta E_n$ associated with the nth cyclotron resonance described by (1):

$$\Delta E_n = 0, \quad n - 1/4 \leq x \leq n, \quad \Delta E_n = \omega(1 - n/x), \quad n \leq x \leq n + 1/4,$$
$$\Delta E_n = \omega[(n + 1/2)/x - 1], \quad n + 1/4 \leq x \leq n + 1/2, \quad \Delta E_n = 0, \quad n + 1/2 \leq x \leq n + 3/4$$

where negative values of $\Delta E_n$ are here replaced by zero. The periodicity in these equations arises from the separation of the orbital energy into its transverse (cyclotron) and longitudinal (free) components relative to the applied magnetic field; in the high-mobility limit, this separation may be asymptotically exact. One can define a planar velocity $\mathbf{v}_c$ by $\Delta E_n = m^* v_{cn}^2 /2$. The interaction of this configurational velocity with $\mathbf{B}$ defines a (steady-state) drift velocity $\mathbf{v}_{Dn} = (\mathbf{v}_{cn} \times \mathbf{B})\tau_b/m^*$, where $\tau_b$ is the mobility relaxation time. Finally, the microwave velocity $\mathbf{v}_w$ is defined by $m^* v_w^2/2 = aP\tau_b$, where P is the microwave power and a is related to the number of carriers in the sample. The phases of these velocities in the (x,y) plane are random.

How will such a many-electron glass respond to the microwave field $\mathbf{F}_W$? In first order there are no net correlations, $<\mathbf{v}_D \cdot \mathbf{F}_W> = 0$. However, in higher orders the carriers can exploit inhomogeneities in the magnetic field to form a variety of open orbits[5]. Experiment in lower mobility samples and perturbation theory have both shown that the magnetoresistance associated with Landau orbitals in magnetic field gradients is always positive[6] to first order in the gradients. It is possible, however, that in very high mobility samples exposed to MW the carriers might organize to form filaments that resemble charge density waves sliding in nearly independent impurity-avoiding[7] channels or wave guides. These filaments are formed because the system can lower its free energy by optimally screening the MW field at long times. In very high quality samples of $NbSe_3$, which has a chain-like crystal structure analogous to self-organized filaments and thus is especially favorable for SCDW, two effects appear[8] in a strong dc electric field **F**: longitudinal ordering (longer longitudinal correlation length for $I/I_c > 1.5$), and transverse ordering (channel narrowing or reduced zigzaging) for $I/I_c > 2.5-3$. Here $I_c$ is the critical current above which depinning occurs by a high-field avalanche mechanism. Although current flows mainly along the pre-formed crystalline chains, it flows in bunches and so the nanoscopic transverse channel ordering is observable.

There is one very favorable feature that differentiates self-organized filaments in a high-mobility 2DEG from SCDW in the $NbSe_3$ case: in the latter, there is always some background resistance associated with parts of the Fermi surface that do not contain a Peierls CDW gap, so zero SCDW resistivity cannot be observed. In the 2DEG in a large **B** the only highly conductive current-carrying parts of the system are the self-organized quantum percolation paths, and so long as these form non-interacting channels, analogous to single-mode optical fibers, the conditions are ideal for ZRS. Of course, the formation of such specific low-entropy states requires a supply of surplus configurational free energy, and it is the latter that accounts for the observed periodicities[2].

Now what happens when a dc electric field **F** in is applied to a high-mobility 2DEG in a crossed magnetic field **B** and a microwave bath? Again in first order there are no correlations, $< \mathbf{v}_W \cdot \mathbf{F} > = 0$, but again in higher orders the mesoscopic CDW filaments pre-formed by the applied microwave field hierarchically organize to screen **F**. These higher order effects can be represented classically by an effective activation energy that is given to lowest order by

$$\Delta E_{an} = <(\mathbf{v}_{Dn} \cdot \mathbf{v}_W)^2> \tau_b (m^*)^2 \qquad (2)$$

Physically this activation energy is similar to a pinning energy for a charge density wave. It can be rewritten in terms of the microwave power P as

$$\Delta E_{an} = P \Delta E_n / P_{0n} \qquad (3)$$

where $P_{0n}$ is a parameter that describes all the configurational factors that are involved in organizing the *n*th CDW. This organization occurs through photonic scattering events, so that quantum mechanically $P_{0n}$ depends on the number of photons or on the MW frequency.

The form of eqn. (3) is unexpected, because it is surprising to see a macroscopic activation energy that is kinetically proportional to a mesoscopic activation energy $\Delta E_n$ through the microwave power P. Such kinetic memory effects are common in spin glasses, and they have recently been described[9] as "hysteretic optimization". The latter is 5-10 times faster (more efficient) in short-range spin glasses than conventional simulated annealing[10]. [Note that all such glassy topological problems are "NP complete", which means that they are analytically insoluble and numerically very difficult, like the traveling salesman problem.] In the presence of long-range electric fields, glassy self-organization can produce much larger effects[8]. The lifetimes $\tau_{SdH}$ = 2.5 ps, $\tau_{CR}$ = 13 ps, and $\tau_t = \mu m^*/e$ = 115 ps, where $\mu$ is the mobility, reported in the earliest experiment[1] are consistent with diffusion and motional narrowing[8], as a simple model of diffusion based on projection operators[11] predicts $\tau_{SdH} \tau_t = \tau_{CR}^2$.

Channels organize independently or in parallel. Moreover, the cyclotron resonance channels associated with closed orbits are inductive, while the open orbits are resistive.

Thus the macroscopic resistivity $R_{xx}(B,\omega)$ is represented by an equivalent circuit composed of independent mesoscopic resistances and inductances

$$R_{xx}(B,\omega) = R/[1+(R/\omega L)^2]^{1/2} \qquad (4)$$

$$R^{-1}(x,P) = \Sigma(R_n)^{-1} \qquad (5)$$

$$R_n(x,P) = R_n(0)\exp[-(P\,\Delta E_n)/(P_{n1}T)] \qquad (6)$$

$$L^{-1}(x,P) = \Sigma(L_n)^{-1} \qquad (7)$$

$$L_n(x,P) = f(P)L_0\,\omega_{cn}^2\tau_n^2/[f(P_{n2})\,(1 + (x-n)^2\omega_{cn}^2\tau_n^2\,] \qquad (8)$$

where there are many adjustable constants assuring dimensional consistency. Here the power P enters R only linearly. In any self-organizing process there will be saturation effects: P can then be replaced by $P_Z \tanh(P/P_Z)$, where $Z = P_{n1}$ or $P_{n2}$. By contrast, the inductive cyclotron orbits pre-exist and $f(P) = aP^{1/2}$ in lower mobility samples[12], but f also saturates at higher P. The square root/linear power dependencies of L and R can be understood if the closed/open orbits are regarded as first- and second-order effects of interactions of the carriers with the applied MW and dc electric fields. These equations can be used to interpret data and study separate resistive and inductive trends as functions of sample mobilities.

Several models have discussed the theoretical aspects of motional ordering of SCDW[8]. The simplest idea is a single-particle one: at large velocities there is motional narrowing of the interactions with impurities. A more sophisticated approach, and more relevant to our present approach to Landau systems, is topological, and predicts two transitions, a plastic flow at intermediate fields, and a dynamic one at larger fields; both of these appear to have been observed in the transverse ordering. The noise spectrum of the SCDW is also best understood topologically in terms of weakly interacting channels[8].

The separation of the measured impedance into non-interacting mesoscopic L and R components may be quite general, as it corresponds to separating paths into sets of open and closed orbits. The latter circle extrema in the landscape potential, while the former avoid them. The velocities of the latter satisfy $\nabla \times \mathbf{v} = 0$, while the former satisfy $\nabla \cdot \mathbf{v} = 0$. Quantum simulations to identify this separation appear to be prohibitively difficult. However, molecular dynamics simulations of a deeply supercooled (glassy) two-component liquid near probe impurities have revealed that the hopping motions that relax applied stress are of two types, translational hopping of clusters ($\nabla \times \mathbf{v} = 0$), or ring-like tunnel motion (momentum circulation, $\nabla \cdot \mathbf{v} = 0$)[13].

The self-organizing effects discussed here seem to be different from those usually considered in the context of self-organized criticality and negative temperatures[14], because here the cyclotron orbits are absent at high T and zero power, and become organized at low T only when irradiated. Thus the self-organization is less like a classical disconnected critical "sand pile" avalanche, and more like quantum constructive SCDW multi-channel formation.

Activation energies $T_0 \sim 20K$ are required in high-mobility quantum wells to erase these channels[3]. These energies are remarkably large compared to the microwave energy • ~ 3K and the Landau energy $•_c$ ~ 2K. They can arise from loss of open-orbit screening energies when the phase coherence of the channels is interrupted at spacings that are 10 times the channel widths. These interruptions apparently occur less often in higher mobility samples, which suggests that they are nucleated by fluctuations in the potential landscape[15,16].

The central concept of this paper is that self-organization is the key to understanding semiclassical, low-field ZRS. Many readers familiar only with the literature of quantum electronic transport may be unfamiliar with this concept, but something very special is required to explain exponentially small resistance. Ordinarily, one uses this phrase in biophysical contexts[4], but it is also useful in many disordered inorganic systems. Note that in the biophysical context there are generally two kinds of functional scaling. Most biological functions are performed by structures that scale geometrically with mass (such as visual cortical functions in primates[17]), but metabolic networks exhibit optimized allometric scaling involving surface/volume ratios in spaces of higher dimensionality d* = d + 1 (warm-blooded animals) or 2d (cold-blooded)[4]. The higher dimensionality is the signature of self-organization, and it cannot be generated perturbatively. Living metabolic networks exhibit many other singular topological properties[18]. Here the metabolic role is played by the microwave field, as it provides a steady supply of low-entropy (long wave length) free energy that is utilized by the sample to form self-organized open orbits for SCDW. The higher dimensionality is reflected in the presence of parallel resistive and inductive electrical multi-channels in eqns. (4)-(8).

There is one other analogy that warrants discussion, and that is the one with the intermediate topological filamentary phase that has been postulated not only as an explanation for high temperature superconductivity in the cuprates, but also as an explanation for the reversibility window in many network glasses[19-21]. Both motionally ordered SCDW in crystals and ZRS states in 2DEG fit in very nicely with the filamentary model for HTSC; in particular, these two examples show that while superconductivity can enhance filamentary self-organization, it is not necessary to achieve it. Thus filaments can be formed at high temperatures (as high as 700K in the cuprates) by self-organization of the dopant configuration, and this accounts for the normal-state transport anomalies and pseudogap formation[19].

In conclusion, the glassy character of disordered impurities prevents us from reaching an exact analytic solution to the problem of the origin of "zero resistance" (exponentially small resistance) states. The self-organized open orbits have much in common with SCDW, but because they arise from screening of the long wave length microwave field, they could also be described as (curvilinear) magnetoplasmons. However, because of their glassy nature, it is not possible to describe these states in terms of the interactions of translationally invariant plasmons with a homogeneous magnetic field, or with impurities by perturbations on Landau levels[5,22]. Instead a more phenomenological approach based on multiple resistive and inductive screening channels appears to be fully adequate to describe the experimental data.

I am grateful to M.A. Zudov for valuable suggestions, and to R.R. Du and R. G. Mani for helpful discussions.

*Postscript.* Several recent preprints have discussed resistance oscillations generated by Landau oscillations in the density of states[23-27]. The resistance oscillations are *supposed* to be large enough to produce negative resistance, implying edge currents or even voltage amplification with a more elaborate circuit geometry. According to the present model, there is no negative resistance, only an exponentially small positive bulk resistance produced by high-field self-organized screening by open and closed orbits that generates separate resistive and inductive mesoscopic currents either avoiding or around landscape extrema.